\documentclass[10pt,preprint]{aastex}

\newcommand{\latin}[1]{{#1}}

\newcommand{\eg}{\latin{e.g.}}
\newcommand{\cf}{\latin{c.f.}}
\newcommand{\Sersic}{S\'ersic}

\newcommand{\avg}[1]{{\langle{#1}\rangle}}
\newcommand{\Avg}[1]{{\left\langle{#1}\right\rangle}}
\def\simless{\mathbin{\lower 3pt\hbox
	{$\,\rlap{\raise 5pt\hbox{$\char'074$}}\mathchar"7218\,$}}} 
\def\simgreat{\mathbin{\lower 3pt\hbox
	{$\,\rlap{\raise 5pt\hbox{$\char'076$}}\mathchar"7218\,$}}} 

\newcommand{\band}[2]{\ensuremath{^{#1}\!{#2}}}

\newcounter{thefigs}
\newcommand{\fignum}{\arabic{thefigs}}

\newcounter{thetabs}

\newcounter{address}

\slugcomment{Submitted to \apj}

\shortauthors{Blanton {\it et al.} (2002)}
\shorttitle{Galaxy dynamics and environment}

\begin{document}
 

\title{Relationship between environment and the broad-band optical
	properties of galaxies in the SDSS\altaffilmark{1}}




\author{
Michael R. Blanton\altaffilmark{\ref{NYU}},
Daniel Eisenstein\altaffilmark{\ref{Steward}},
David W. Hogg\altaffilmark{\ref{NYU}},
David J. Schlegel\altaffilmark{\ref{Princeton}}, and
J. Brinkmann\altaffilmark{\ref{APO}}
}

\altaffiltext{1}{Based on observations obtained with the
Sloan Digital Sky Survey\label{SDSS}} 
\setcounter{address}{2}
\altaffiltext{\theaddress}{
\stepcounter{address}
Center for Cosmology and Particle Physics, Department of Physics, New
York University, 4 Washington Place, New
York, NY 10003
\label{NYU}}
\altaffiltext{\theaddress}{
\stepcounter{address}
Steward Observatory, 933 N. Cherry Ave., Tucson, AZ 85721
\label{Steward}}
\altaffiltext{\theaddress}{
\stepcounter{address}
Princeton University Observatory, Princeton,
NJ 08544
\label{Princeton}}
\altaffiltext{\theaddress}{
\stepcounter{address}
Apache Point Observatory,
2001 Apache Point Road,
P.O. Box 59, Sunspot, NM 88349-0059
\label{APO}}

\begin{abstract}
We examine the relationship between environment and the luminosities,
surface brightnesses, colors, and profile shapes of luminous galaxies
in the Sloan Digital Sky Survey (SDSS).  For the SDSS sample, galaxy
color is the galaxy property most predictive of the local
environment. Galaxy color and luminosity jointly comprise the most
predictive pair of properties. At fixed luminosity and color, density
is not closely related to surface brightness or to \Sersic\ index ---
the parameter in this study that astronomers most often associate with
morphology. In the text, we discuss what measureable residual
relationships exist, generally finding that at red colors and fixed
luminosity, the mean density decreases at the highest surface
brightnesses and \Sersic\ indices. In general, these results suggest
that the morphological properties of galaxies are less closely related
to galaxy environment than are their masses and star-formation
histories.
\end{abstract}


\section{Motivation}
\label{motivation}

While galaxy formation theory has had remarkable success in predicting
certain properties of galaxies (their clustering, for example), it
does not yet successfully predict the detailed joint distribution of
galaxy properties, such as their luminosities, colors, surface
brightnesses, and profile shapes. Thus, we do not have a complete
understanding of the physical processes associated with galaxy
formation, such as gas infall, disk dynamics, galaxy mergers, star
formation, and feedback from supernovae and central black holes.
However, from observations we do know that galaxy properties correlate
with galaxy environment and therefore that one physical parameter of
importance is the local density. A more detailed understanding of the
relationship of galaxy properties to that physical parameter may
therefore shed light on galaxy formation.

Much work on this subject focuses on the relationship between galaxy
morphology and environment (e.g. \citealt{hubble36a, oemler74a,
dressler80a}, or the more recent work of
\citealt{hermit96a,guzzo97a,giuricin01a}). These works all find that
earlier type (elliptical) galaxies are more strongly clustered than
later type (spiral) galaxies. Another approach is to consider
clustering as a function of more objective (though not necessarily
more relevant) quantities such as spectral type (\citealt{norberg02a})
or photometric color, surface brightness, luminosity, or profile shape
(e.g. \citealt{hashimoto99a,zehavi02a}). Since all the properties
mentioned above (morphology, spectroscopic properties, and photometric
properties) are highly correlated, it is not surprising that
clustering is a function of all of them. The question naturally arises
as to which properties are correlated with environment independently
of the others. \citet{norberg02a}, \citet{zehavi02a},
\citet{budavari03a}, and \citet{hogg03b} have begun this process by
measuring the clustering of galaxies as a function of luminosity and
other properties jointly.

In this paper we systematically explore the local environments of
luminous galaxies as a function of their colors, luminosities, surface
brightnesses and radial profile shapes, using the Sloan Digital Sky
Survey (SDSS; \citealt{york00a}). Where necessary, we have assumed
cosmological parameters $\Omega_0 = 0.3$, $\Omega_\Lambda = 0.7$, and
$H_0 = 100 h$ km s$^{-1}$ Mpc$^{-1}$.

\section{Data}
\label{data}

The SDSS is taking $ugriz$ CCD imaging of $10^4~\mathrm{deg^2}$ of the
Northern Galactic sky, and, from that imaging, selecting $10^6$
targets for spectroscopy, most of them galaxies with
$r<17.77~\mathrm{mag}$ \citep[\eg,][]{gunn98a,york00a,abazajian03a}.
Automated software performs all of the data processing: astrometry
\citep{pier03a}; source identification, deblending and photometry
\citep{lupton01a}; photometricity determination \citep{hogg01a};
calibration \citep{fukugita96a,smith02a}; spectroscopic target
selection \citep{eisenstein01a,strauss02a,richards02a}; spectroscopic
fiber placement \citep{blanton03a}; and spectroscopic data reduction.
An automated pipeline measures the redshifts and classifies the
reduced spectra, modeling each galaxy spectrum as a linear combination
of stellar populations (Schlegel et al., in preparation).

The sample is statistically complete, with small incompletenesses
coming primarily from (1) galaxies missed because of mechanical
spectrograph constraints \citep[6~percent;][]{blanton03a}, which does
lead to a slight under-representation of high-density regions, and (2)
spectra in which the redshift is either incorrect or impossible to
determine ($<1$~percent).  In addition, there are some galaxies ($\sim
1$~percent) blotted out by bright Galactic stars, but this
incompleteness should be uncorrelated with galaxy properties.

Galaxy luminosities and colors \citep[measured by the standard SDSS
petrosian technique;][]{petrosian76a,strauss02a} are computed in fixed
bandpasses, using Galactic extinction corrections \citep{schlegel98a}
and $K$-corrections \citep[computed with \texttt{kcorrect
v1\_16};][]{blanton03b}.  They are $K$-corrected not to the redshift
$z=0$ observed bandpasses but to bluer bandpasses $\band{0.1}g$,
$\band{0.1}r$ and $\band{0.1}i$ ``made'' by shifting the SDSS $g$,
$r$, and $i$ bandpasses to shorter wavelengths by a factor of 1.1
\citep[\cf,][]{blanton03b, blanton03d}.  This means that galaxies at
exactly redshift $z=0.1$ (typical of the SDSS sample used here) all
have the trivial $K$-correction $K(z=0.1) = -2.5 \log_{10} (1.1)$
regardless of their spectral energy distribution. 

To determine the galaxy profile shape, we fit the radial galaxy
profile using a simple \Sersic\ measurement (\citealt{sersic68a}), accounting
for seeing. The form of the \Sersic\ profile is:
\begin{equation}
\label{sersic}
I(r) = A \exp\left[ - (r/r_0)^{1/n} \right].
\end{equation}
From a fit of the radial profile to this form, we derive a \Sersic\
index $n$ (which takes a value of 1 for exponential galaxies and 4 for
de Vaucouleurs galaxies), half-light sizes $r_{50}$ in kpc, and
surface-brightnesses of galaxies $\mu_{50}$. As described in the
Appendix, our \Sersic\ measurements here are an updated version of
those in \citet{blanton03d}, for which there was a bias in the
\Sersic\ indices of concentrated galaxies. From tests of simulated
data processed with the SDSS photometric pipelines plus our \Sersic\
fitting procedure, we find that our \Sersic\ indices are
underestimates at high \Sersic\ index (we measure about $n=3.5$ for
input galaxies with $n=4$).

For the purposes of computing large-scale structure statistics, we
have assembled a subsample of SDSS galaxies known as the NYU LSS
\texttt{sample10}. This sample is a superset of the recent Data
Release One (\citealt{abazajian03a}).  For each galaxy in
\texttt{sample10}, we have computed a volume $V_\mathrm{max}$
representing the total volume of the Universe (in
$h^{-3}~\mathrm{Mpc^3}$) in which the galaxy could have resided and
still made it into the sample.  The calculation of these volumes is
described elsewhere \citep{blanton03d}.

We estimate the overdensity on $1\,h^{-1}~\mathrm{Mpc}$ scales from a
deprojected angular correlation function.  Around each spectroscopic
target galaxy, galaxies are counted in the SDSS imaging in the
magnitude range corresponding to $M^\ast\pm1~\mathrm{mag}$
(passively-evolved and $K$-corrected as for an early-type galaxy) and
within $5h^{-1}~\mathrm{Mpc}$ \citep[transverse, proper;
\eg,][]{hogg99cosm} at the spectroscopic galaxy redshift.  We weight
the count to recover the estimated overdensity averaged over a
\emph{spherical} three-dimensional Gaussian window $e^{-r^2/2a^2}$
with a radius of $a=1h^{-1}~\mathrm{Mpc}$
(proper). \citet{eisenstein03a} give the details of the weighting and
the method for correcting for the survey mask.  The results do not
depend on an assumed model of the correlation function but do depend
inversely on the normalization of the luminosity function at the
redshift in question.  One advantage of this method is that the
density can be estimated with a volume-limited and yet reasonably
dense set of galaxies, even at the furthest reaches of the
spectroscopic catalog.  Another advantage is that the estimator is not
affected by redshift distortions.  If the spatial correlation function
$\xi(r)$ has the form $r^{-\gamma}$, then the mean overdensity around
galaxies $\left<\delta_1\right>$ is
$(2/\sqrt{\pi})\,\Gamma([3-\gamma]/2)\,\xi(1\,h^{-1}~\mathrm{Mpc})$, 
where $\Gamma(x)$ is the standard gamma function.

In what follows, the mean overdensities of a range of galaxy
subsamples are presented. The mean has the advantage that it recovers
a true three-dimensional overdensity, smoothed with a well-defined
filter. While it is also interesting to study the properties of
galaxies selected to be in environments of different overdensity (as
in, {\it e.g.} Hogg et al. in preparation), the overdensity estimators used
here have signal-to-noise too low for use in galaxy selection. This
density estimator is generally {\it only} informative in the mean.

We selected the sample of galaxies used here to have apparent
magnitude in the range $14.5<r<17.77~\mathrm{mag}$, redshift in the
range $0.05<z<0.22$, fixed-frame absolute magnitude in the range
$M_{^{0.1}i}>-24.0~\mathrm{mag}$, and enough surrounding sky area to
reliably measure overdensities.  These cuts left 114994 galaxies.

\section{Dependence of overdensity on galaxy properties}
\label{results}

\subsection{What pair of galaxy properties is most closely related to
	overdensity?}

In this section we demonstrate that the pair of galaxy properties
(among the four we consider here) most predictive of local overdensity
is color and luminosity.

\begin{figure}
\figurenum{\fignum}
\plotone{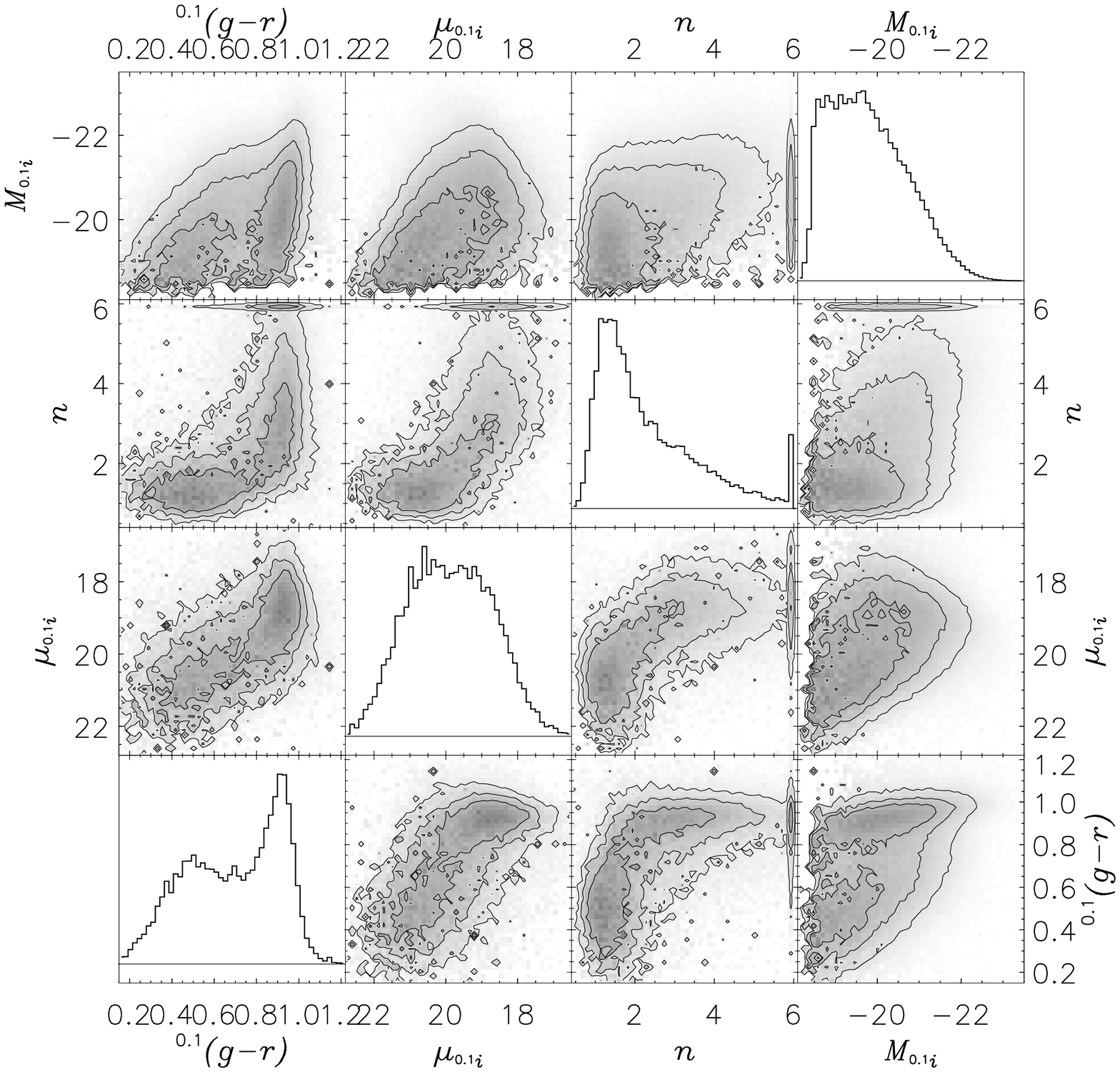}
\caption{\label{bid_greyscale} Bivariate number density distributions
(that is, $1/V_{\mathrm{max}}$ weighted) of properties of galaxies in
our sample. All off-diagonal images have a square root
stretch. Contours indicate the regions containing 52\%, 84\%, and 97\%
of the number density. The upper and lower triangles are mirror images
of each other. The histograms along the diagonal show the distribution
of galaxies with each property on a linear scale. }
\end{figure}

Figure \ref{bid_greyscale} shows for our sample the
$1/V_{\mathrm{max}}$-weighted distribution of $\band{0.1}{(g-r)}$
color, half-light surface brightness $\mu_{\band{0.1}{i}}$, \Sersic\
index $n$, and absolute magnitude $M_{\band{0.1}{i}}$.  The diagonal
plots show the distribution of each property individually as a
histogram on a linear scale; the off-diagonal plots show it as a
two-dimensional histogram expressed as a grey scale. The contours on
the greyscale indicate the areas containing 97\%, 84\% and 52\% of the
distribution (from the outer contours to the inner contours). This
figure repeats the results of \citet{blanton03c} but for the more
restrictive set of luminosities we consider here, and as discussed
above including improved estimates of the \Sersic\ indices. A small
number of galaxies ($< 1\%$) have a best-fit \Sersic\ index $n=6$,
corresponding to the largest value we allow.

\begin{figure}
\figurenum{\fignum}
\epsscale{1.091}
\plotone{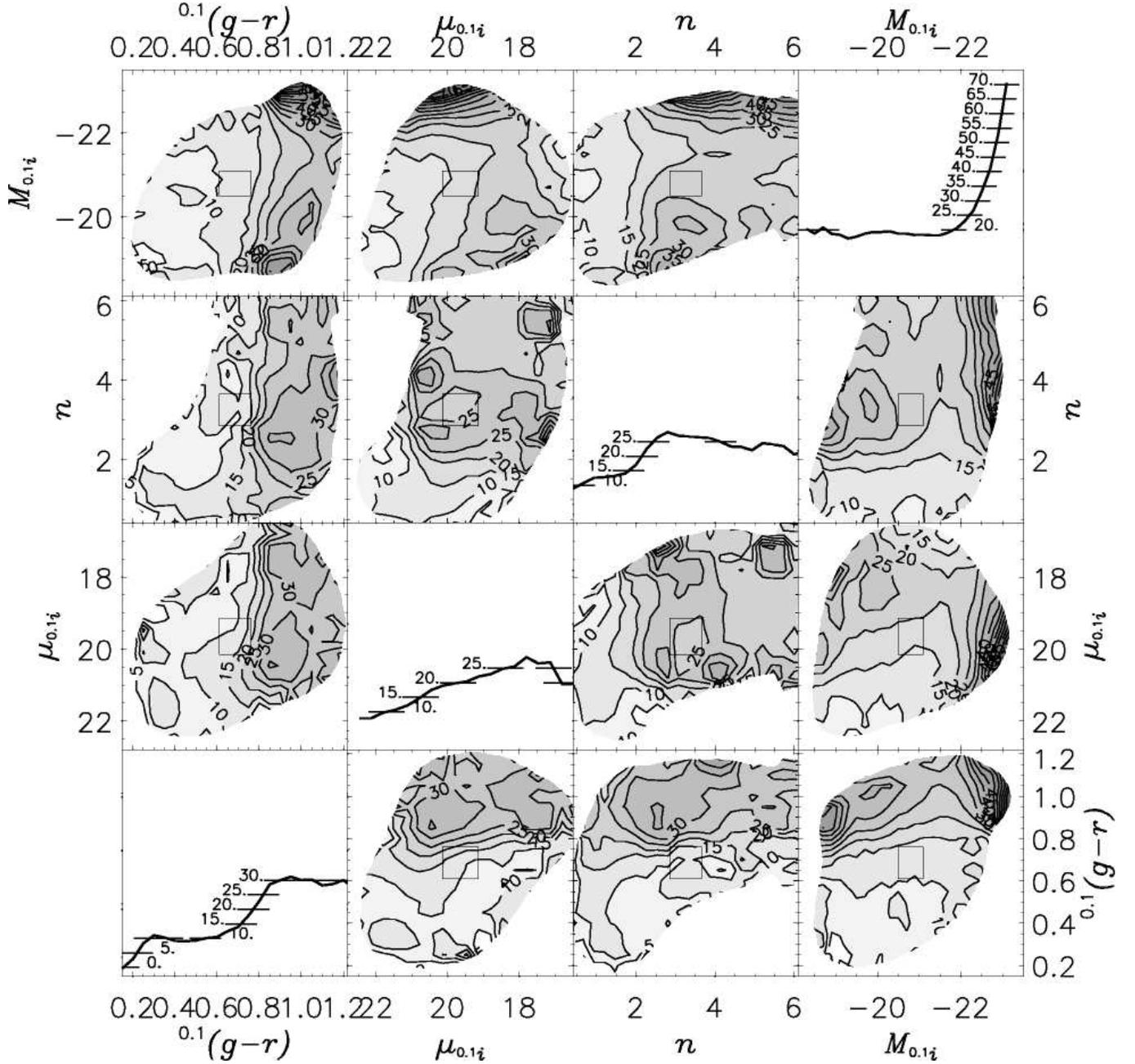} 
\epsscale{1.0}
\caption{\label{bid} Mean local overdensity as a function of pairs of
galaxy properties. Off-diagonals show the mean overdensity as a
color-coded contour plot in which darker areas indicate galaxies in
denser environments. For example, in the lower right corner, the
blue, low luminosity galaxies are on average in the least dense
environments and red, high luminosity galaxies are on average in the
most dense environments. Plots along the diagonal show the mean
overdensity as a function of each property on a linear
scale. Labeled cross bars indicate where the mean crosses various
thresholds. }
\end{figure}

Figure \ref{bid} shows a contour plot of the mean overdensity around
galaxies as a function of each property and each pair of
properties. The mean is calculated in a sliding box with width
$(0.15,1.0,0.8,0.6)$ in color, surface brightness, \Sersic\ index, and
absolute magnitude respectively. The diagonal panels show the plot
using a one-dimensional analog of the contour plot; the lines indicate
the mean and the labeled cross bars (designed to be the width of the
sliding box) indicate where the mean crosses various thresholds. In
the off diagonal plots, regions colored dark indicate higher
density. If the sliding box contains fewer than 200 galaxies, the
result is ignored and colored pure white.

\begin{figure}
\figurenum{\fignum}
\epsscale{1.091}
\plotone{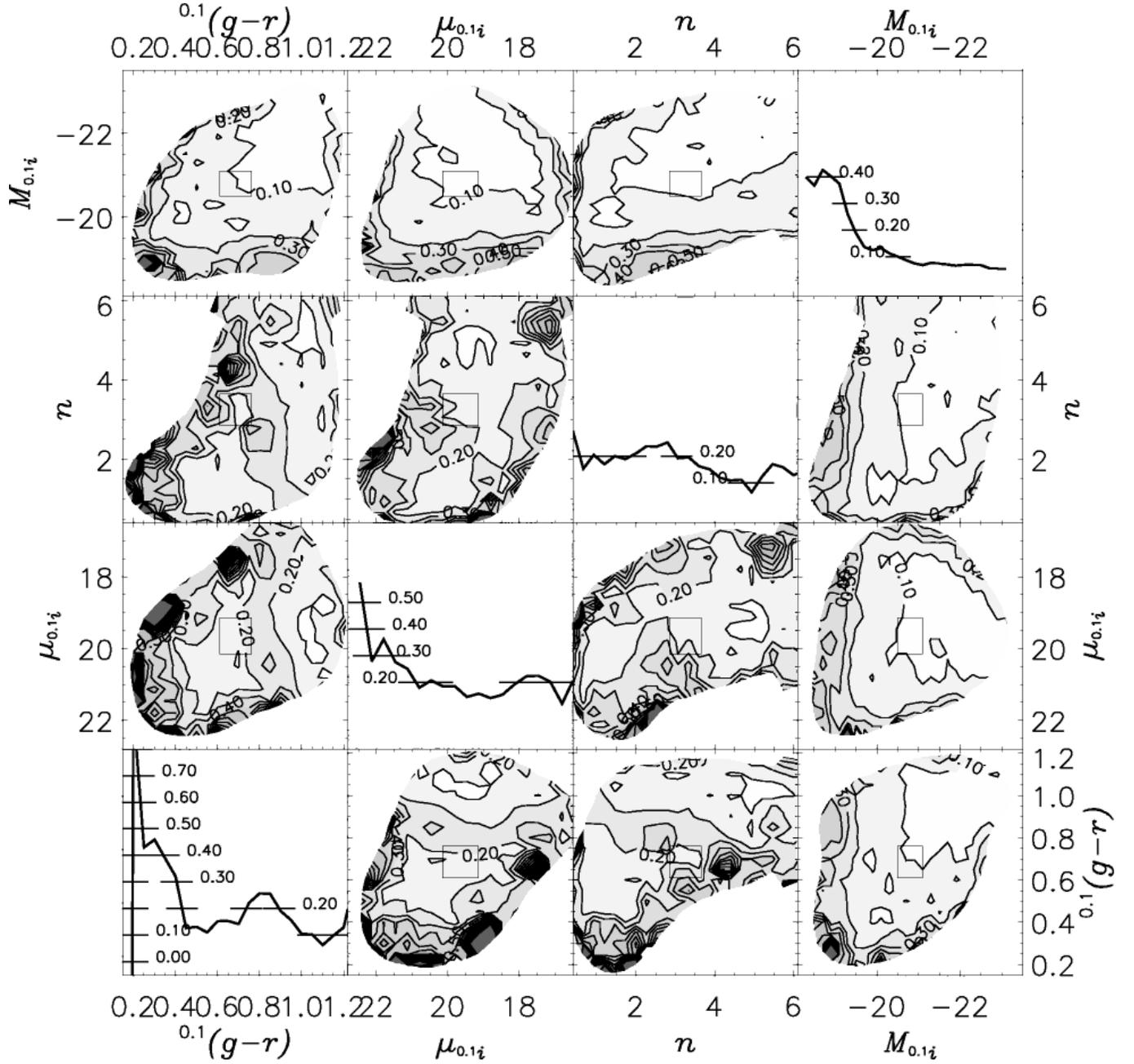}
\epsscale{1.0}
\caption{\label{bid_fracerr} Fractional uncertainties (from
jackknife resampling) in the average density as a function of pairs
of galaxy properties.}
\end{figure}

We evaluate the uncertainties in these results using the jackknife
technique, dividing the survey area into ten roughly contiguous,
roughly equal area sections. We then recalculate our results ten
times, each time leaving out one section.  The jackknife estimate of
the standard deviation in the mean overdensity is the square root of
the second moment of this set of results around the original result
(\citealt{lupton93a}):
\begin{equation}
\Avg{\Delta x^2}^{1/2} = \left[\frac{N-1}{N} \sum_{i} (x_i - {\bar
		x}_i)^2 \right]^{1/2}
\end{equation}
This method thus includes the effects of cosmic variance in the
uncertainty budget.  Figure \ref{bid_fracerr} shows the standard
deviation as a fraction of the mean overdensity calculated in this
way. Over most of the plots the uncertainty is of order 10--30\%.

What properties or pair of properties are most closely related to
overdensity? This question is not completely well-defined. For
example, even if the dependence of mean overdensity on galaxy
properties were purely linear, the slopes of those linear relations
would be impossible to compare in a meaningful way --- what relates
the units of \Sersic\ index $n$ to those of absolute magnitude $M$?
One way of expressing the question that is at least independent of
metric choices is to ask what property or pair of properties is most
predictive of an object's local density. That is, take the number
density weighted variance of the $\delta_1$ measurements
$\sigma^2\equiv \avg{(\delta_{1,i} - {{\bar \delta}_1})^2}$. Now
measure ${\bar \delta}_1(X,Y, \ldots)$ as a function of one or more
galaxy properties, as in Figure \ref{bid}. It must be the case that
\begin{equation}
\sigma_{X}^2 \equiv \Avg{(\delta_{1,i}-{\bar \delta}_1(X_i))^2} \le \sigma^2 
\end{equation}
As stated above, this statistic is independent of the units of the
quantities $X$. It {\it is} dependent on one's choice of binning for
these quantities, but the binning we use is reasonable in the sense
that each bin contains a large number of galaxies, is somewhat bigger
than the errors in each parameter, and is smaller than any features in
the actual distribution of the given parameter.  Note that when we
perform this calculation, we in fact weight each galaxy by
$1/V_{\mathrm{max}}$; weighting by number density rather than
luminosity density (or some other weighting) is an arbitrary
choice. What parameter $X$ or pair of parameters $X$ and $Y$,
minimizes this variance? Table \ref{xytable} lists the values of
$\sigma_{XY}-\sigma^2$ for each pair $X$ and $Y$.

For galaxies in the range of luminosities considered here,
$\band{0.1}{(g-r)}$ color is the most predictive of overdensity in
this sense. In Figure \ref{bid}, the dependence on luminosity appears
more impressive, but this dependence only affects a small fraction of
the total number density of galaxies in this sample.  

The most predictive pair of properties is $\band{0.1}{(g-r)}$ and
$M_{\band{0.1}{i}}$ taken together.  The dependence of overdensity on
color and luminosity takes an interesting form, as noted previously by
\citet{hogg03b}. At high luminosity clustering is a strong function of
luminosity. At low luminosity it is a strong function of color. Red
galaxies around $M_{\band{0.1}{i}\ast} \sim -21$ are less clustered in
general than red galaxies of higher or lower luminosity.

\begin{figure}
\figurenum{\fignum}
\epsscale{1.091}
\plotone{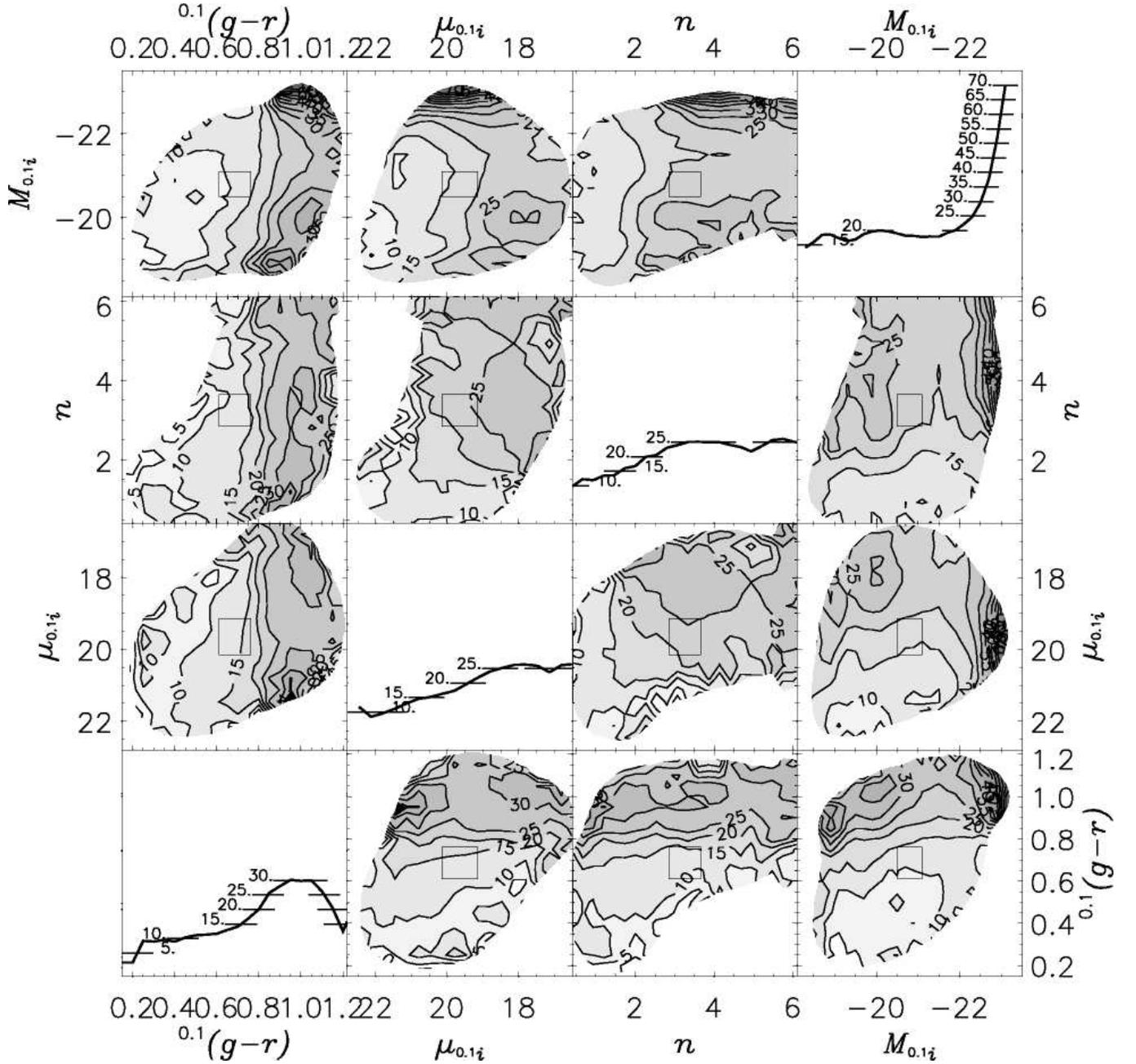}
\epsscale{1.0}
\caption{\label{fake_bid} Same as Figure \ref{bid}, but instead of
using the actual overdensity of each galaxy, we replace it with the
mean overdensity for a galaxy of that color and absolute
magnitude. Thus, this figure reproduces the color and absolute
magnitude dependence (slightly smeared) by design but not
necessarily the other dependencies. Comparing this figure with
Figure \ref{bid} reveals that the dependence on surface brightness
and concentration at low luminosity is at least partly a by-product
of the dependence on color at those luminosities, but that the
dependence on surface brightness and concentration at high
luminosity is not.}
\end{figure}

\begin{figure}
\figurenum{\fignum}
\epsscale{1.091}
\plotone{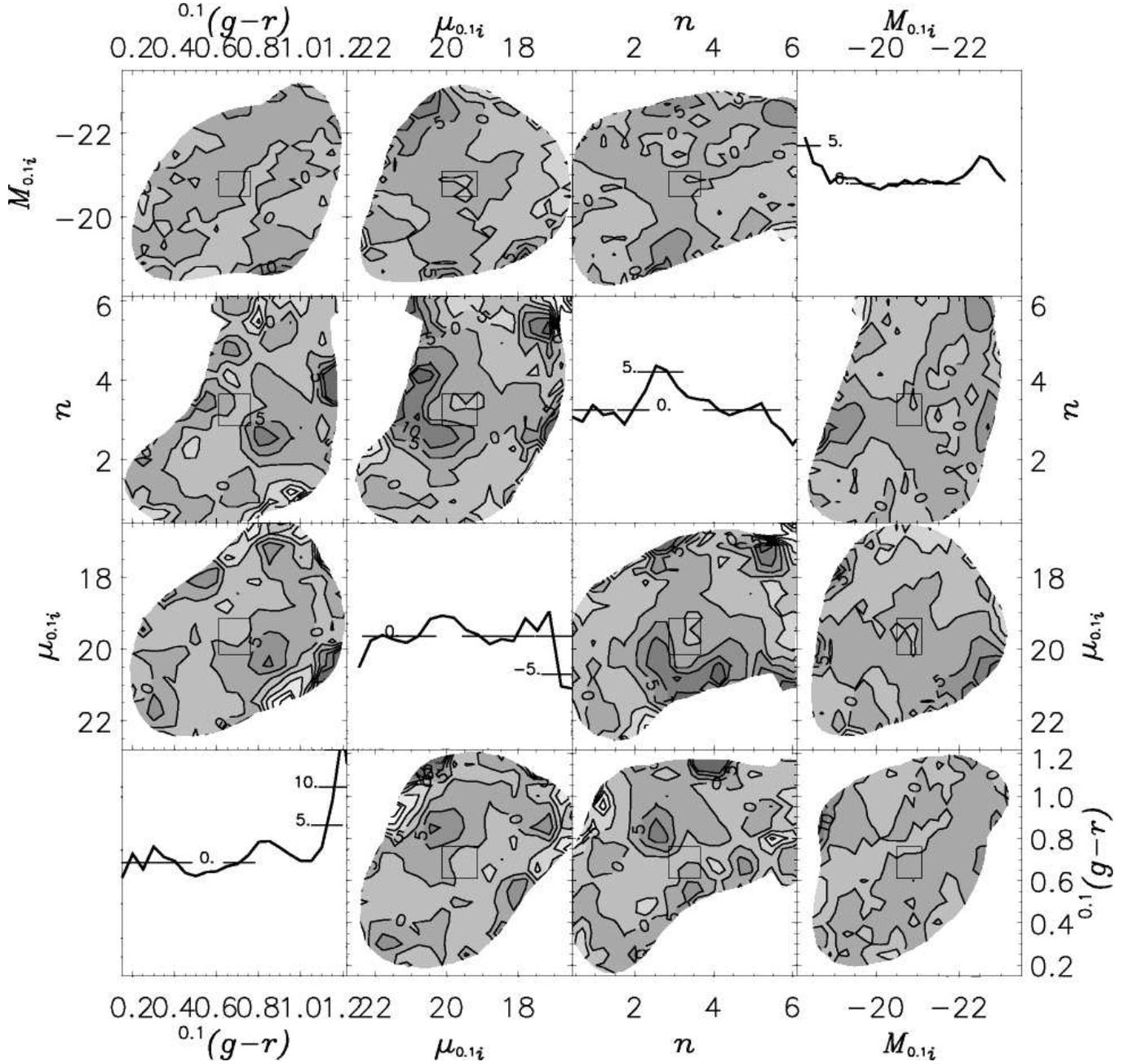}
\epsscale{1.0}
\caption{\label{diff_bid} The difference between Figures \ref{bid} and
  \ref{fake_bid}, shown in the same manner as those figures.}
\end{figure}

\Sersic\ index $n$ and surface brightness $\mu_{\band{0.1}{i}}$ are
both correlated with luminosity and color. Therefore, one should ask
whether the remaining dependence of mean overdensity on these two
parameters in Figure \ref{bid} are really independent of the
dependence on luminosity and color or whether we can express the
former as a consequence of the latter. To do so, we assign each galaxy
in our sample a ``fake'' overdensity based on the mean overdensity and
its variance given its luminosity and color, using the lower right
panel of Figure \ref{bid}. In Figure \ref{fake_bid} we show the
dependence of the mean of this fake overdensity as a function of all
parameters. The form of the lower right panel is reproduced, although
the dependence is slightly smeared because of the finite resolution of
the grid we use to assign the fake overdensity. Figure \ref{diff_bid}
shows the difference between the actual mean overdensity in Figure
\ref{bid} and the ``fake'' mean overdensity in Figure \ref{fake_bid}.

From Figures \ref{fake_bid} and \ref{diff_bid} we see that we can
express some of the dependence of overdensity on \Sersic\ index and
surface brightness in terms of the dependence on color and
luminosity. For example, at low luminosity there is a dependence of
overdensity on increasing \Sersic\ index and increasing surface
brightness apparent in Figure \ref{fake_bid}, indicating that the
similar dependence seen in Figure \ref{bid} is not independent of
color and luminosity.

However, Figures \ref{fake_bid} and \ref{diff_bid} make it clear that
some features of Figure \ref{bid} {\it are} independent of color and
luminosity. At high luminosity, overdensity increases in Figure
\ref{bid} towards lower surface brightness while overdensity remains
constant in Figure \ref{fake_bid}. In Figure \ref{diff_bid}, this
result appears as an increase in density towards low surface
brightness at high luminosity.

We have looked at corresponding plots using all the other pairs of
properties on which to base the ``fake'' overdensities. Unless
luminosity is included explicitly as a member of the pair, the strong
luminosity dependence for galaxies more luminous than $L_\ast$ and the
lack of dependence on luminosity for less luminous galaxies cannot be
replicated. Thus, this luminosity dependence is fundamental --- not a
derivative of the dependence on any other property. The combination of
surface brightness and luminosity fails to reproduce the color and
\Sersic\ index dependence or to reproduce the high overdensities of
the red, low luminosity galaxies. The combination of \Sersic\ index
and luminosity reproduces the dependence on surface brightness but
fails to reproduce the strong dependence of density on color or
(again) the high overdensities of red, low luminosity galaxies.

\subsection{What is the residual dependence on the other properties?}

The previous subsection shows that color and luminosity are most
predictive of a galaxy's overdensity. However, it also suggests that
there is a residual dependence on \Sersic\ index and surface
brightness. Here we explore this residual dependence in more detail.  

\begin{figure}
\figurenum{\fignum}
\plotone{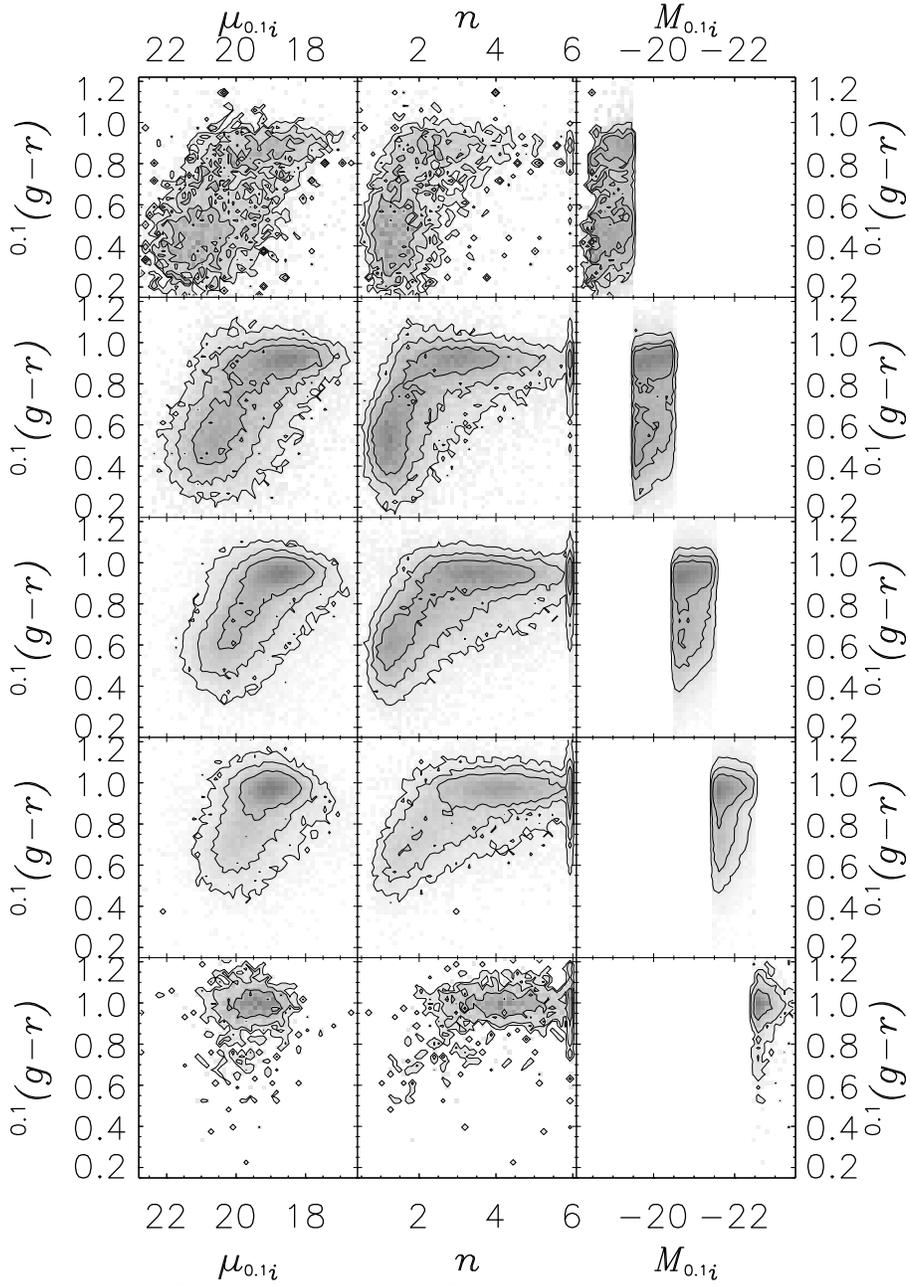}
\caption{\label{bidld_greyscale} Similar to Figure \ref{bid_greyscale},
	for five ranges of absolute magnitude, as shown. Here we show only
	the number density as a function of color and other properties.}
\end{figure}

\begin{figure}
\figurenum{\fignum}
\epsscale{0.7244}
\plotone{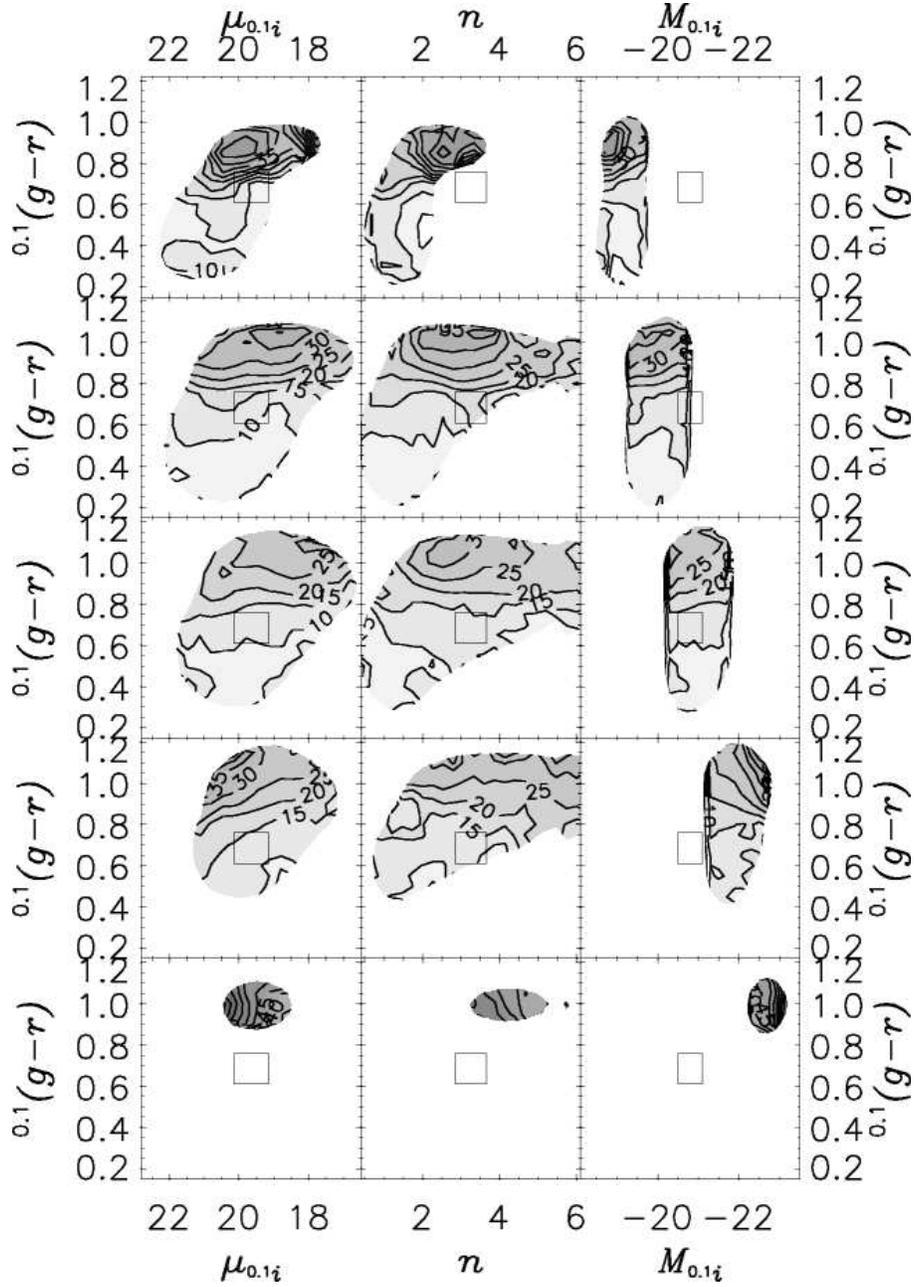}
\epsscale{1.0}
\caption{\label{bidld} Similar to Figure \ref{bid}, for the five
	ranges of absolute magnitude used in Figure
	\ref{bidld_greyscale}. At fixed color and luminosity there exists
	dependencies on surface brightness and concentration. In particular,
	at high luminosity there is a strong dependence of clustering on
	surface brightness and concentration, in the sense that lower
	surface brightness and lower \Sersic\ index galaxies are in denser
	regions. At lower luminosity, the higher surface brightness galaxies
	lower \Sersic\ index galaxies are in denser regions. }
\end{figure}

\begin{figure}
\figurenum{\fignum}
\epsscale{0.7244}
\plotone{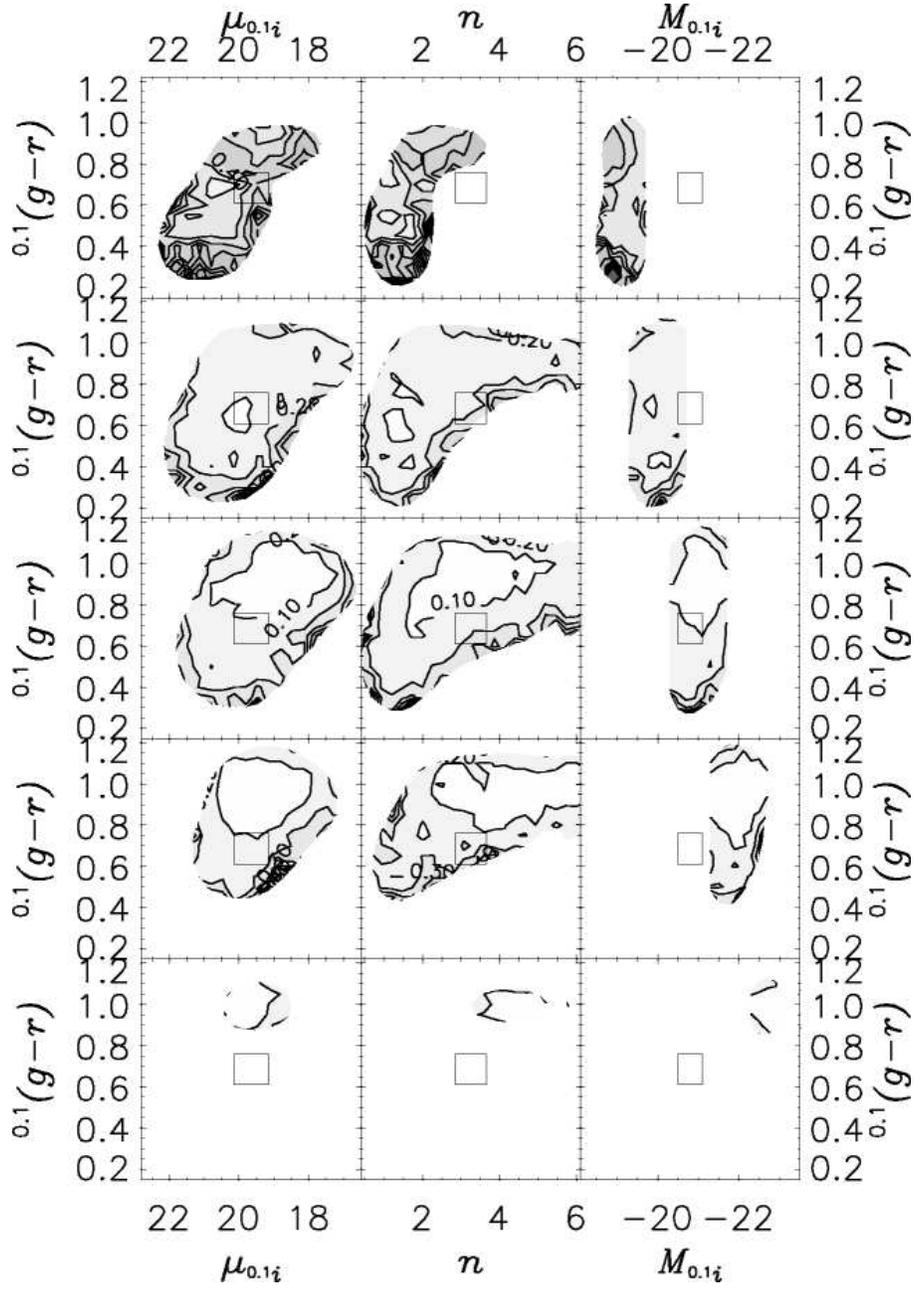}
\epsscale{1.0}
\caption{\label{bidld_fracerr} Similar to Figure \ref{bid_fracerr},
	showing fractional uncertainties in the mean overdensities plotted
	in Figure \ref{bidld}.}
\end{figure}

Figure \ref{bidld_greyscale} shows the number density distribution of
galaxies as a function of color, surface brightness, and \Sersic\
index, for five ranges of absolute magnitude (analogously to Figure
\ref{bid_greyscale}). Figure \ref{bidld} shows the mean local
overdensity for these sets of galaxies (analogously to Figure
\ref{bid}). The right column consists simply of slices of the lower
right panel of Figure \ref{bid}. However, the left two columns now
show the dependence of overdensity on color and surface brightness and
\Sersic\ index for fixed absolute magnitude. In Figure
\ref{bidld_fracerr} we show the fractional standard deviation
estimated from jackknife (analogous to Figure \ref{bid_fracerr}).

At high luminosity, there is dependence of overdensity on surface
brightness and \Sersic\ index --- lower surface brightness and less
concentrated galaxies are in denser regions. The surface brightness
dependence persists to luminosities just above $L_\ast$ and exists
even for blue ($\band{0.1}{(g-r)} \sim 0.6$) galaxies.

At luminosities less than $L_\ast$ there is no dependence on surface
brightness for blue galaxies but for red galaxies higher surface
brightness galaxies exist in denser environments. Meanwhile, there is
also little dependence on \Sersic\ index for blue galaxies of these
luminosities, but for red galaxies a middle value $n=3$ of the
\Sersic\ index indicates denser regions. 

In order to test the significance of this
effect, we selected galaxies with $\band{0.1}{(g-r)} > 0.9$, $-21.8 <
M_i < -20.2$, and $2 < n < 4$, regressed the overdensities of these
galaxies against \Sersic\ index $n$, and evaluated the uncertainties
in the measured slope using the jackknife samples. We found that in
this range $\delta_1 \approx 41 - 4.3 n$, where the 1$\sigma$
uncertainties on the slope were $\sigma \sim 1.1$. Thus, this result
appears significant. On the other hand, it is worth noting the caveat
that this statistical measurement is {\it a posteriori} selected as
interesting, out of a large number of possible measurements.

\begin{figure}
\figurenum{\fignum}
\epsscale{0.7244}
\plotone{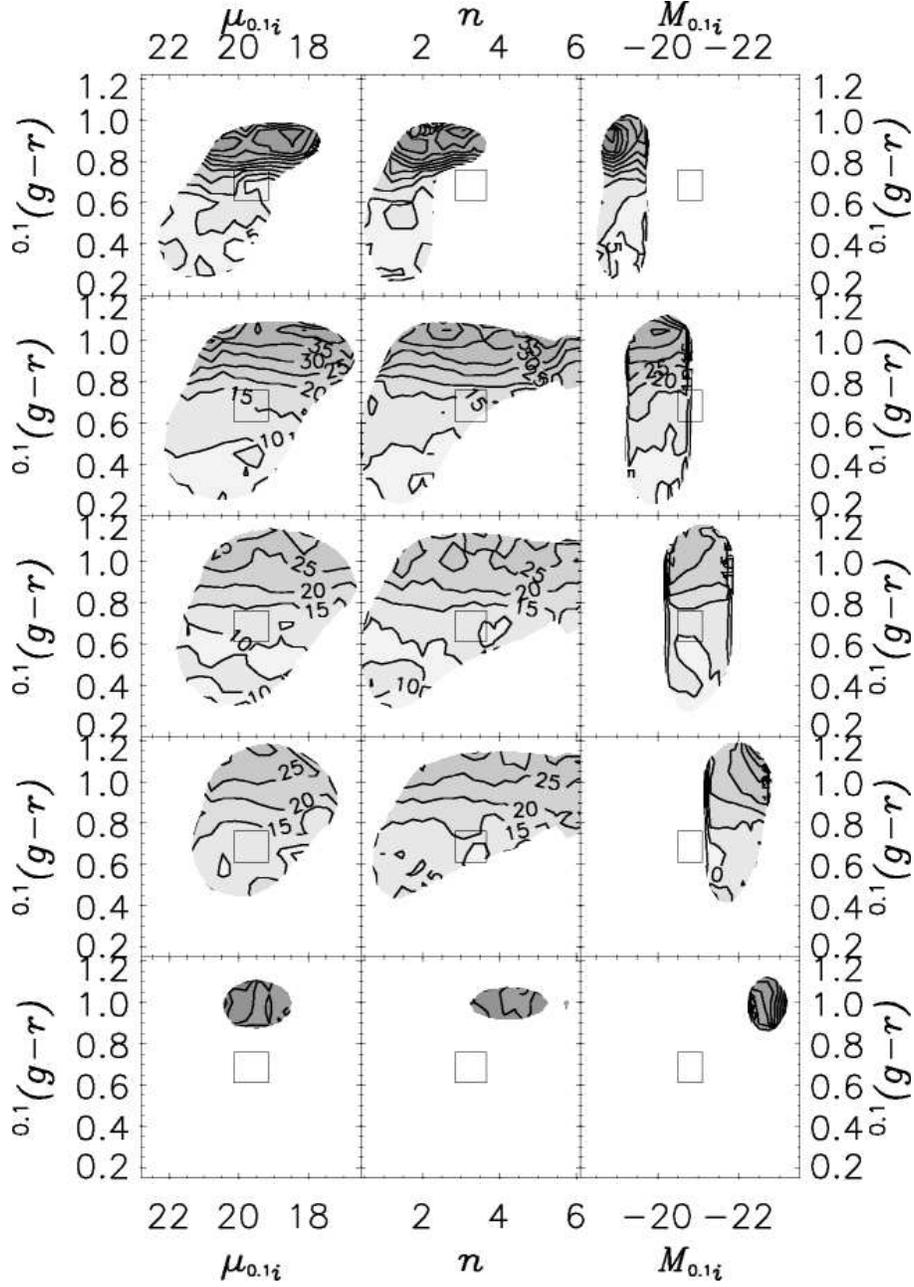}
\epsscale{1.0}
\caption{\label{fake_bidld} Similar to Figure \ref{fake_bid}, showing
	the same plot as Figure \ref{bidld} but replacing the overdensity
	for each galaxy with the mean overdensity for its absolute magnitude
	and color. The correlations between density and luminosity within
	each luminosity slice clearly do not result in artificial features
	in the dependence of density on \Sersic\ index and surface
	brightness. }
\end{figure}

One might worry that because the slices in absolute magnitude in
Figure \ref{bidld} are not infinitely thin, some of the dependence on
\Sersic\ index and surface brightness in Figure \ref{bidld} is simply
a result of the absolute magnitude dependence combined with the
correlation of absolute magnitude with \Sersic\ index and surface
brightness. Using the same technique as for Figure \ref{fake_bid}, we
show in Figure \ref{fake_bidld} the results using the fake
overdensities based on the color and absolute magnitude of each
galaxy. It is clear from this figure that little of the dependence in
Figure \ref{bidld} results from this effect. In addition, we have
repeated the test described in the last paragraph using narrower
ranges of absolute magnitude and found consistent results.

\section{Summary and Discussion}
\label{discussion}

We have found here that the dependence of environment on color and
luminosity has a nontrivial form and captures much of the interesting
dependence of environment on galaxy properties. In particular, the
measurement we make which is most closely related to morphology, the
\Sersic\ index, appears to be less related to environment than is
color. This result does not exclude the possibility that more detailed
measurements of morphology may have a closer link with environment.

The form of our measured dependence of density on galaxy properties is
likely related to phenomena already noted in the literature. The
preponderance of giant ellipticals in clusters (\citealt{dressler80a})
is clearly related to the strong dependence on luminosity for red,
luminous galaxies. The denseness of the environments of the lower
luminosity red galaxies may be a signature of the luminous end of the
dwarf elliptical population which also reside in clusters
(\citealt{ferguson89a,depropris95a,mobasher03a}). The strong
dependence of environment on color for bluer galaxies is clearly
related to the lack of star forming galaxies in clusters. In short,
these results are clearly related in some way to the classical
density-morphology relation. However, in our results the parameter
most closely related to classical morphology (\Sersic\ index $n$)
appears to have a relatively weak relationship with density
independent of color and luminosity.

While color and luminosity are the properties most closely related to
overdensity, there are {\it some} residual dependences with respect to
surface brightness and \Sersic\ index. In particular, at high
luminosity lower \Sersic\ index and lower surface brightness galaxies
are more strongly clustered. At least part of this effect must be
related to the existence of the cD galaxies at the centers of
clusters, known to be less concentrated and lower surface brightness
than typical giant ellipticals (\citealt{schombert86a}). However, the
dependence on surface brightness persists even for blue galaxies,
which cannot be caused by classical cD galaxies.

At lower luminosity, less concentrated red galaxies are more highly
clustered than more concentrated red galaxies. We do not believe that
this effect is related to any previously described morphology-density
relationship. Nor do we believe that this result contradicts any
previous investigations, none of which accounted for the
interdependence of galaxy properties as systematically and completely
as we have here. It is worth noting that it is {\it possible} that our
photometric measurements of the \Sersic\ profile are being affected by
nearby neighbors in such a way as to make galaxies with many neighbors
appear less concentrated than the average red galaxy.

\section{Future Work}
\label{future}

The measurements we make here may be improved in two significant
ways. First, we can make more detailed morphological measurements. We
are currently working on more detailed morphological measurements for
a nearby set of the galaxies.  Second, we can push the sample to lower
luminosity. Both of these efforts require considering a lower redshift
sample. Unfortunately, in the current configuration of the SDSS survey
it is difficult to evaluate the environments of galaxies at low
redshift because of the edge effects of the survey. Nevertheless, with
the Northern Galactic Cap section of the SDSS being steadily filled
in, we will have a much larger contiguous data set to work on.

Here we have considered the environments of galaxies as a function of
their properties, because the environmental measures we use are low
signal-to-noise. However, we are working in parallel on environmental
measures which are higher precision and looking at the dependence of
galaxy properties as a function of their environment
(Hogg et al. in preparation).

Finally, as acknowledged in Section \ref{discussion}, it is possible
that our photometric measurements have systematic biases which are a
function of environment. Although we do not believe that these effects
have greatly affected our main result here that density is not
strongly dependent on \Sersic\ index at fixed luminosity and color, we
will be checking for these biases as part of a larger program of
testing the photometric results of the SDSS using simulated data.

\acknowledgments

We are indebted to Robert H. Lupton for help in performing the
simulated data tests in the Appendix.  We would like to thank David
Weinberg and Michael Strauss for useful comments and discussion. The
development and debugging of {\tt sample10} would have been impossible
without Andreas Berlind, Chiaki Hikage, Nikhil Padmanabhan, Adrian
Pope, Shiyin Shen, Yasushi Suto, Max Tegmark and Idit Zehavi.  Thanks
to Sam Roweis for interesting discussions on data modelling.  This
work would not have been possible without the tremendous {\tt
idlutils} library of software developed by Doug Finkbeiner, Scott
Burles, and DJS, and the Goddard distribution of software distributed
by Wayne Landsman. We only wish there were publications to cite for
these tools. MB and DWH acknowledge NASA NAG5-11669 for partial
support.  DJE is supported by NSF grant AST-0098577 and by a Alfred
P. Sloan Research Fellowship. MB is grateful for the hospitality of
the Department of Physics and Astronomy at the State University of New
York at Stony Brook, who kindly provided computing facilities on his
frequent visits there.

Funding for the creation and distribution of the SDSS has been
provided by the Alfred P. Sloan Foundation, the Participating
Institutions, the National Aeronautics and Space Administration, the
National Science Foundation, the U.S. Department of Energy, the
Japanese Monbukagakusho, and the Max Planck Society. The SDSS Web site
is {\tt http://www.sdss.org/}.

The SDSS is managed by the Astrophysical Research Consortium (ARC) for
the Participating Institutions. The Participating Institutions are The
University of Chicago, Fermilab, the Institute for Advanced Study, the
Japan Participation Group, The Johns Hopkins University, Los Alamos
National Laboratory, the Max-Planck-Institute for Astronomy (MPIA),
the Max-Planck-Institute for Astrophysics (MPA), New Mexico State
University, University of Pittsburgh, Princeton University, the United
States Naval Observatory, and the University of Washington.

\appendix

\section{Tests of the \Sersic\ Measurements}

The \Sersic\ model fits we use here are different than those that
\citet{blanton03c} use. We have slightly improved the \Sersic\ fitting
code and removed a major bug. Here we describe in full the fitting
method, its performance on simulated SDSS data, and the differences
from the previous version.

First, we fit a model consisting of three axisymmetric gaussians to a
31x31 pixel (non-axisymmetric) image of the seeing at the center of
each field determined by the SDSS Postage Stamp Pipeline (PSP; described in
\citealt{stoughton02a}). We equal weight the pixels in the fit and
minimize the squared residuals with respect to the model parameters
using the IDL routine {\tt mpfitfun}, an implementation of the
Levenberg-Marquardt method written by C. Markwardt. The radial profile
of the fit is typically within a fraction $10^{-3}$ of the actual
radial profile of the seeing image at all radii. 

For each galaxy, we fit the axisymmetric \Sersic\ model of
Equation (\ref{sersic}) to the mean fluxes in annuli output by the
SDSS photometric pipeline {\tt photo} in the quantities {\tt profMean}
and {\tt profErr} (\citealt{stoughton02a} list the radii of these
annuli). {\tt photo} outputs these quantities only out to the annulus
which extends beyond twice the Petrosian radius, or to the first
negative value of the mean flux, whichever is largest. In any case, we
never consider data past the 12th annulus, whose outer radius is
$68.3''$. We define:
\begin{equation}
\chi^2 = \sum_i [ (\mathtt{profMean}_i -
  \mathtt{sersicMean}_i(A,n,r0) 
) / \mathtt{profErr}_i]^2
\end{equation}
where {\tt sersicMean}$_i(A,n,r0)$ is the mean flux in annulus $i$ for
the \Sersic\ model convolved with the three-gaussian seeing model for
the given field. 
We do not perform this convolution on the pixel grid,
so while the seeing model includes the pixel convolution, it does not
include it exactly.  In practice we evaluate model annuli on a large
grid of \Sersic\ radii, \Sersic\ indices, and Gaussian seeing widths,
and when fitting interpolate off the grid. We again use {\tt mpfitfun}
for the minimization.

We have evaluated the performance of this algorithm in the following
way. Taking a sampling of the parameters of our fits from the Main
galaxy sample, we have generated about 1200 axisymmetric fake galaxy
images, which we refer to as ``fake stamps.'' There are some
subtleties to producing correct fake stamps, in particular with
integrating the central few pixels. We first evaluate the \Sersic\
profile at the center of each pixel; then we sort the pixels by that
intensity, and perform a full integration for the 36 pixels with the
largest fluxes.

In order to simulate the performance of {\tt photo}, we have
distributed the fake stamps among SDSS fields. For each band, we
convert the fake stamps to SDSS raw data units, convolve with the
estimated seeing from PSP, and add Poisson noise using the estimates
of the gain. We add the resulting image to the SDSS raw data at a
random location on the frame, including the tiny effects of
nonlinearity in the response and the less tiny flat-field variation as
a function of column on the chip. We run {\tt photo} on the resulting
set of images to extract and measure objects and then run our
\Sersic\ fitting code. This procedure thus includes the effects of
seeing, noise, and sky subtraction. We have tested that our results
remain the same if we insert images using an alternative estimate of
the seeing based simply on stacking nearby stars (still fitting using
our three-gaussian fit to the PSP seeing estimate).

\begin{figure}
\figurenum{\fignum}
\plotone{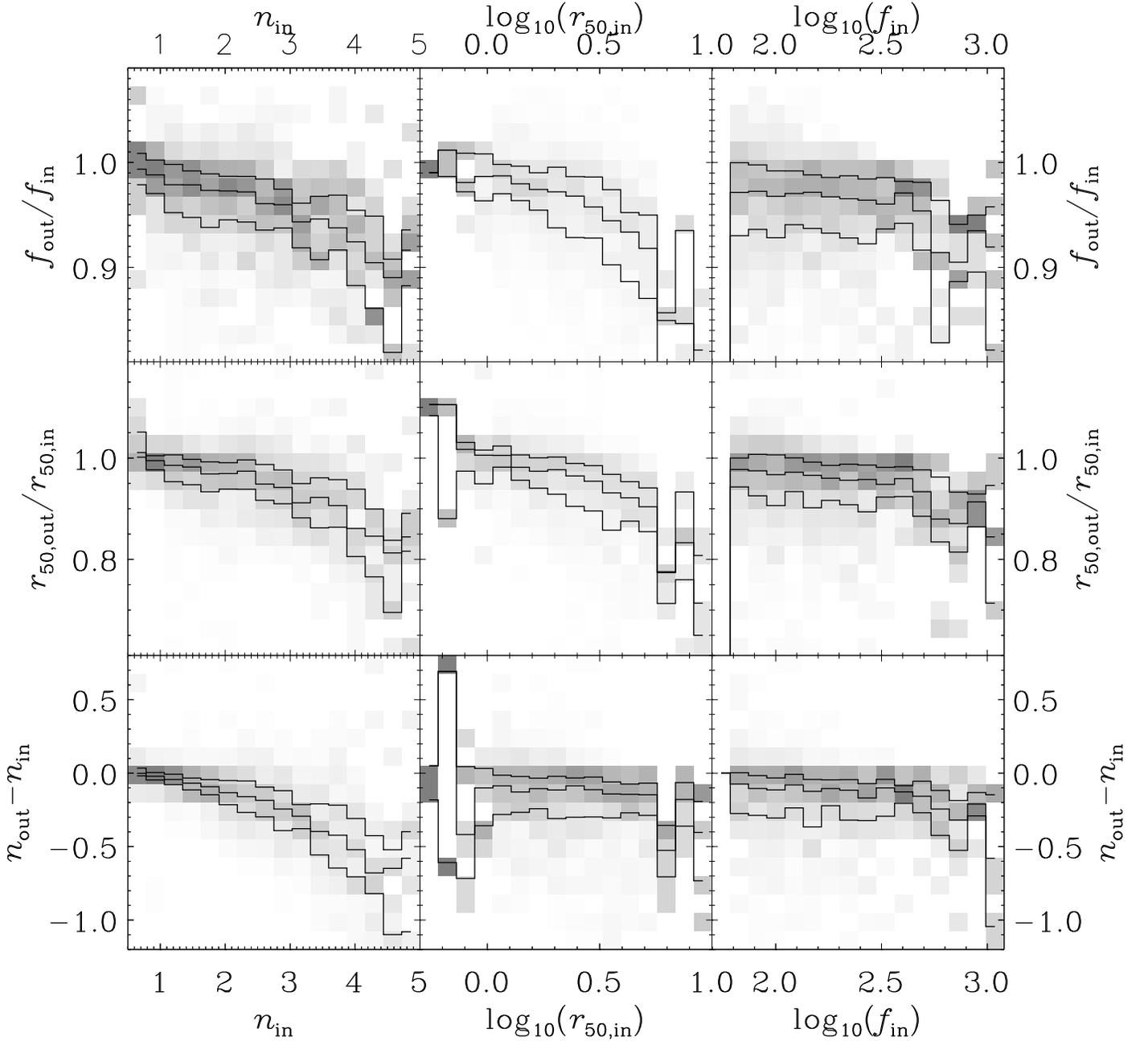}
\epsscale{1.0}
\caption{\label{sersic_compare} Residuals of the fit \Sersic\
  parameters as a function of the input \Sersic\ parameters for a set
  of 1200 simulated galaxies inserted into raw data and processed with
  the SDSS photometric pipeline plus the \Sersic\ fitting procedure
  in Appendix A. The fluxes and sizes are those associated with the
  \Sersic\ fit. The greyscale represents the conditional probability
  of the $y$-axis measurement given the $x$-axis input; the lines show
  the quartiles of that distribution.}
\end{figure}

Figure \ref{sersic_compare} displays the distribution of fit
parameters in the $r$-band (converting $A$ and $r_0$ to total flux $f$
and half-light radius of the profile $r_{50}$), as a function of the
input parameters. Each panel shows the conditional distribution of the
quantity on the $y$-axis as a function of quantity on the
$x$-axis. The fluxes $f$ are expressed in nanomaggies, such that
$f=100$ corresponds to $m = 17.5$, near the flux limit of the Main
galaxy sample (the general conversion of nanomaggies to magnitudes is
$m=22.5-2.5\log_{10}(f)$). The lines show the quartiles of the
distribution. At all \Sersic\ indices, sizes, and fluxes, the
performance is good.

For larger sizes, sizes and fluxes are underestimated by about 10\%
and 15\% respectively, while the \Sersic\ index is constant over a
factor of ten in size. For high \Sersic\ indices, the sizes and fluxes
are slightly underestimated (again by about 10\% and 15\%) while the
\Sersic\ index itself is underestimated by (typically) -0.5 for a de
Vaucouleurs galaxy --- meaning that a true de Vaucouleurs galaxy
yields $n\sim 3.5$ in our fits.  This remaining bias is not much
larger than the uncertainty itself and is comparable to the bias one
expects (in the opposite direction) from neglecting non-axisymmetry.

The bias is partly due to our approximate treatment of the seeing, but
mostly due to small errors in the sky level (at the level of 1\% or
less of the sky surface brightness) determined by the photometric
software. If one fits for the sky level, one can recover the \Sersic\
indices (and fluxes and sizes) of the fake galaxies far more
accurately. However, because the \Sersic\ model is not a good model
for galaxy profiles, for real data the fits apply unrealistically high
changes to the sky level to attain slight decreases in $\chi^2$. The
resulting sizes and fluxes of the largest and brightest galaxies are
obviously wrong. Thus, we satisfy ourselves that the measurements we
obtain with the fixed sky level yield approximately the right answer
for galaxies which are actually \Sersic\ models, and for other
galaxies merely supply a seeing-corrected estimate of size and
concentration.

When we ran this test on the original version of our code that
\citet{blanton03c} used, we found that the \Sersic\ fits to
de~Vaucouleurs galaxies were underestimating the \Sersic\ index by
about 1.3. This error occurred partly because we incorrectly
interpreted parameters for a double Gaussian fit to the seeing
produced by PSP, and somewhat less importantly, because we were not
fitting explicitly for the sky level. These errors caused (relative to
the fits using the new version) an underestimate in the \Sersic\ index
for concentrated galaxies by about 0.8, and an overestimate (by up to
40\%) of the galaxy size for small galaxies. Relative to the new
version, the flux of the old version was underestimated by only a tiny
amount ($\sim 2\%$) for concentrated galaxies and hardly at all for
exponential galaxies. The old version still measured $n=1$ galaxies
correctly, and for higher \Sersic\ indices had the property that
measured \Sersic\ indices ran monotonically with the concentration of
the galaxies.

\newpage

\clearpage
\clearpage

\setcounter{thefigs}{0}

\clearpage
\stepcounter{thefigs}

\clearpage
\stepcounter{thefigs}

\clearpage
\stepcounter{thefigs}

\clearpage
\stepcounter{thefigs}

\clearpage
\stepcounter{thefigs}

\clearpage
\stepcounter{thefigs}

\clearpage
\stepcounter{thefigs}

\clearpage
\stepcounter{thefigs}

\clearpage
\stepcounter{thefigs}

\clearpage
\stepcounter{thefigs}

\newpage
\clearpage
 
\begin{deluxetable}{ccccccc}
\tablewidth{0pt}
\tablecolumns{7}
\tablecaption{\label{xytable} Explanatory power of pairs of galaxy
	properties}
\tablehead{ &  & \quad & & Property Y & & \cr 
Property X & $\sigma_X^2-\sigma^2$ & & \band{0.1}{(g-r)} & 
$\mu_{\band{0.1}{i}}$ & $n$
	& $M_{\band{0.1}{i}}$ }
\startdata
\band{0.1}{(g-r)} & 
-82.0
&
 & 
 \nodata 
&
-88.8
&
-93.3
&
-116.2
\cr
$\mu_{\band{0.1}{i}}$ & 
-23.8
&
 & 
-88.8
&
 \nodata 
&
-48.4
&
-46.6
\cr
$n$ & 
-37.0
&
 & 
-93.3
&
-48.4
&
 \nodata 
&
-55.3
\cr
$M_{\band{0.1}{i}}$ & 
-4.0
&
 & 
-116.2
&
-46.6
&
-55.3
&
 \nodata 
\cr
\enddata
\tablecomments{ The second column gives $\sigma_{X}^2-\sigma^2$, where
$\sigma^2$ is the variance in the estimates of the density and
$\sigma_{X}^2$ is the variance around the mean relations on the
diagonal panels of Figure \ref{bid} for each property $X$. Each entry
of the right four columns is $\sigma_{XY}^2-\sigma^2$, where
$\sigma^2$ is the variance in the estimates of the density and
$\sigma_{XY}^2$ is the variance around the mean relations in Figure
\ref{bid} for each pair of properties $X$ and $Y$. Lower values
indicate that the properties have more explanatory power; that is,
that they are in this sense more related to the local
overdensity. Color is the single most explanatory parameter, while
color and luminosity comprise the most explanatory pair of
parameters. }
\end{deluxetable}

\end{document}